\documentclass[12pt]{iopart}
\usepackage{iopams}
\usepackage{setstack}
\usepackage{amssymb}
\usepackage{rotating}
\usepackage{placeins}
\usepackage{lscape}  
\begin{document}
\vspace{-10ex}
\title{Quasi-Exact Solvability and Deformations of $Sl(2)$ Algebra}
\vspace{-2ex}
\author{Arunesh Roy$^1$, Abhijit Sen$^2$ and Prasanta K. Panigrahi$^{1*}$}
\address{$^1$Department of Physical Sciences, Indian Institute of Science Education and Research -Kolkata, 
Mohanpur Campus, P.O. BCKV Campus Main Office, Mohanpur - 741252, Nadia, West Bengal, India.}
\address{$^2$Suri Vidyasagar College, Suri-731101, West Bengal, India.}
\eads{\mailto{pprasanta@iiserkol.ac.in}$^*$}
\begin{abstract}
Algebraic structure of a class of differential equations including Heun is shown to be related with the deformations of $sl(2)$ algebra.
These include both quadratic and cubic ones. The finite dimensional representation of cubic algebra is explicitly shown to describe a 
quasi-exactly solvable system, not connected with $sl(2)$ symmetry. Known finite dimensional representations of $sl(2)$ emerge under special conditions. 
We answer affirmatively the question raised by Turbiner: 
''Are there quasi-exactly solvable problems which can not be represented 
in terms of $sl(2)$ generators?" and give the explicit deformed symmetry underlying this system. 

\end{abstract}
\vspace{-3ex}
\vspace{-2ex}
\submitto{\JPA}\vspace{-2ex}
\section{Introduction}\vspace{-2ex}
It is well known that, only a few one dimensional Schr\"{o}dinger eigenvalue problems exhibit exact solvability \cite{Flugge}. 
Recently, the class of spectral problems showing quasi-exact solvability has attracted considerable attention \cite{Ushveridge}.
Quasi-exactly solvable (QES) systems have partially tractable energy spectrum. Only a few eigenvalues and their  eigenfunctions 
are analytically approachable \cite{Turbiner, Bender, Atre}. 
They have a deep connection with finite dimensional representations of $sl(2)$ group and are also connected with 
equilibrium electrostatic  configurations \cite{Turbiner}. Interestingly, a quasi-exactly solvable system has been identified, 
which has connection with Heun differential equation \cite{Ronveaux} and is not amenable to the $sl(2)$ based classification \cite{AKhare}. 

In this paper, we analyze the algebraic structure of a wide class of differential equations, including Heun as a subsystem. It is found that the underlying symmetries of
 this class of equations are deformed $sl(2)$. These deformations are of cubic and quadratic type for Heun and confluent or bi-confluent Heun
differential equations, respectively. It is also found that, the finite dimensional representation of cubic algebra, describes a quasi-exactly solvable system, not connected 
with $sl(2)$ symmetry. 

The paper is organized as follows. In the following section, we will consider a class of differential equations with regular singularities ranging from 0 to 3, of which 
Heun is a subclass. We identify the operators in the respective differential equations, from which the deformation of the algebra emerges. We also compute the Casimir characterizing the
representations. Subsequently, we consider the exact and quasi-exact solvability of the differential equations, when related to appropriate eigenvalue equations. 
The respective conditions of exact and quasi-exact solvability produces Hypergeometric and
a type of Heun differential equation. Finite dimensional representation of $sl(2)$ algebra is shown to emerge under certain conditions, 
which  characterize a number of quasi-exactly solvable systems. 
 Our analysis, not only answers positively the question raised by Turbiner \cite{Turbiner} : 
''Are there quasi-exactly solvable problems which can not be represented 
in terms of $sl(2)$ generators?", but also gives the explicit deformed symmetry underlying this system.

\begin{landscape}
\begin{table}\label{eqntable}
\caption{The parameters for the Heun class and the Jacobi differential equations and their related symmetries.}
\vspace{2ex}
\begin{tabular}{ll ccccccccll } 
\hline 
\\[-1ex]
Differential equations \hspace{-10ex} & \multicolumn{9}{c}{Values of parameters $a_{i}(i = 0,\ldots , 8)$ } \hspace{-10ex} & \hspace{-10ex}Symmetry algebra\\ [4ex] 
\hline 
\\[-1ex]
  & $a_0$ & $a_1$ & $a_2$ & $a_3$ & $a_4$ & $a_5$ & $a_6$ & $a_7$ & $a_8$ &  \\ [5ex] 
\hline 
\\[-1ex]
{Heun Equation}  & 1 & $-(c+1)$ & c & 0 & $\gamma+\delta+\varepsilon$ & $-[\gamma(c+1)+\delta c +\varepsilon)]$ & $\gamma c$ & $\alpha\beta$ & $-q$ & $Cubic $\\[6ex]
{Confluent Heun} &0 & $1$ & $-1$ & 0 & $\nu$ & $\gamma+\delta-\nu$ & $-\gamma$ & $\alpha\nu$ &$ -\sigma$ & \hspace{-2ex}$Quadratic$ \\[5ex]
{Bi-Confluent Heun}  &0 & $0$ & 1 & 0 & $-2$ & $-\beta$ & $\alpha+1$ & $\gamma-\alpha-2$ &$ -\frac{1}{2}[\delta+(\alpha+1)\beta]$ & \hspace{-2ex}$Quadratic$ \\[5ex]
{Doubly Confluent}  &0 & $1$ & 0 & 0 & $-1$ & $\tau$ & $\nu$ & $-\alpha$ & $q$ & $Linear$ \\[5ex]
{Jacobi} &$0$ & $-1$ & 0 & 1 & $0$ & $-(\alpha+\beta+2)$ & $\beta-\alpha$ & 0 & $\ n(n+\alpha+\beta+1)$ & $Cubic$\\[5ex]
\hline 
\end{tabular}
\end{table}
\end{landscape}
\FloatBarrier

\section{Heun Class of Differential Equations and Deformed $Sl(2)$ Algebra}\vspace{-2ex}
\label{sec1}
As mentioned before, the spectral Schr\"{o}dinger equation,
\begin{equation}\label{eqS}
\hat{H}\psi=E\psi, \hat{H}= -\frac{d^2}{dx^2}+V(x), \qquad x\in(-\infty,+\infty)\;\mbox{or}\; x\in[0,\infty)
\end{equation}
can be connected with differential equations, having different singularity structure by suitable 
change of variables and appropriate point canonical transformations \cite{Sudarshan}. These differential equations are  generically of the form, 
\begin{equation}\label{Gendiff}
\left[f_{1}(x)\frac{d^2}{dx^2}+f_2(x)\frac{d}{dx}+f_3(x)\right]\psi(x) = 0,
\end{equation}
where 
\begin{numparts}
\begin{eqnarray}
f_{1}(x)=a_{0}x^3+a_{1}x^2+a_{2}x+a_{3},\\
f_{2}(x)=a_{4}x^2+a_{5}x+a_{6},\\
f_{3}(x)=a_{7}x+a_{8};
\end{eqnarray}
\end{numparts}
with $ a_{i} \in \mathbf{R} \ \mbox{for} \ i = 0,\ldots,8 $. 
These equations have regular singularities, 
varying between 0 to 3, depending on the values of $a_{i}$'s.

Keeping in mind quasi-exactly solvable problems and their connection 
with $sl(2)$ algebra, we now start with a finite dimensional representation 
of $sl(2)$ algebra with spin $j$, in a space of monomials $x^{j+m}$  $(m \leq \vert j \vert)$. 
The following generators 
\begin{equation}\label{Generator}
J^{+}= x^2\frac{d}{dx}-2jx,\qquad
J^{0}= x\frac{d}{dx}-j,\qquad
J^{-}= \frac{d}{dx},
\end{equation} 
satisfy the closed algebra
\begin{equation}\label{gensl2}
[J^{+},J^{-}]=-2J^{0} ,\qquad [J^{0},J^{\pm}] = \pm J^{\pm}.
\end{equation} 
The fact that the representation space is finite and algebraically interrelated explains the quasi-exact solvability
 of the corresponding Schr\"{o}dinger equation \cite{Shifman}.

To analyze the symmetry of the QES problems for the aforementioned class of differential 
equations \eref{Gendiff}, we introduce a set of operators $\{P_{+},P_{0},P_{-}\}$ in the space 
of monomials such that :
\begin{equation*}\label{Ps}
P_{+}x^{n} = c_{+}x^{n+1},\qquad
P_{0}x^{n} = c_{0}x^{n} ,\qquad
P_{-}x^{n} = c_{-}x^{n-1}.
\end{equation*}
The general differential equation \eref{Gendiff} can be cast in terms of $\{ P_{+},P_{0},P_{-} \}$ 
operators if $ a_{3}=0 $ : 
\begin{equation}\label{pkpeq}
\left[P_{+}+F(P_{0})+P_{-}\right] \psi (x) = 0.
\end{equation}   
Clearly
\begin{numparts}
\begin{eqnarray}
P_{+} = a_{0}x^3\frac{d^2}{dx^2}+a_{4}x^2\frac{d}{dx}+a_{7}x,\label{Genop1}\\
F(x\frac{d}{dx})=a_{1}x^2\frac{d^2}{dx^2}+a_{5}x\frac{d}{dx}+a_{8},\label{Genop2} \qquad 
P_{0} = x\frac{d}{dx}-j,\\
\mbox{and}\qquad
P_{-}=a_{2}x\frac{d^2}{dx^2}+a_{6}\frac{d}{dx}.\label{Genop3}
\end{eqnarray}
\end{numparts}

$F(x\frac{d}{dx})$ can be simplified in terms of $P_{0}$:
\begin{equation*}
F(P_{0}) = a_{1}P_{0}^2+((2j-1)a_{1}+a_{5})P_{0}+(a_{1}j^{2}-(a_{1}-a_{5})j+a_{8}).
\end{equation*}
The algebraic structure can be inferred from the following closed form, 
\begin{equation}
\left[P_{+},P_{-}\right] = \alpha_{1} P_{0}^3 + \beta_{1} P_{0}^2 + \gamma_{1} P_{0} + \delta_{1}=f(P_{0}) \qquad\mbox{and}
\qquad [P_{0},P_{\pm}]=\pm P_{\pm}
\end{equation}
where,
\begin{eqnarray*}
\alpha_{1}=-4a_{0}a_{2},\qquad\beta_{1}=6(1-2j)a_{0}a_{2}-3a_{2}a_{4}-3a_{0}a_{6} \\
\gamma_{1}=
(-2a_{0}a_{2}+3a_{0}a_{6}-2a_{4}a_{6}-2a_{2}a_{7}+a_{2}a_{4})  
\\+2(6a_{0}a_{2}-3a_{2}a_{4}-3a_{0}a_{6})j+(-12a_{0}a_{2})j^2, \\
\delta_{1}=-a_{6}a_{7}+(-2a_{0}a_{2}+3a_{0}a_{6}-2a_{4}a_{6}-2a_{2}a_{7}+a_{2}a_{4})j 
\\+(6a_{0}a_{2}-3a_{2}a_{4}-3a_{0}a_{6})j^2+(-4a_{0}a_{2})j^3.
\end{eqnarray*}
The eigenvalue of the Casimir operator, $C = J^{-}J^{+}+g(J^{0})$ is $(a_{6}a_{7})$, with $g(J^{0})-g(J^{0}-1)=f(J^{0})$ \cite{Rocek} for the present case. 
Hence, the constituent operators of the differential 
equation \eref{Gendiff} satisfy the \emph{cubic deformation} of the 
$sl(2)$ structure \cite{nonlinalgeb,deformap,deform,Bambah}. Some well known differential equations \cite{Ronveaux,Govindarajan,Lay,Ovsiyuk,Kazakov,Jha}
with the above structure, are listed in Table-1.
   The solutions to the above equations can be obtained employing a recent approach, which connects 
   the space of monomials to the solution space \cite{PKP}. For this purpose, the differential equation 
   should be cast in the form \eref{pkpeq}
 provided the condition $F(x\frac{d}{dx})x^{\lambda}=0$. This leads to,
 \begin{equation}
 \lambda_{\pm}=\frac{1}{2a_{1}}\left[-(a_{5}-a_{1})\pm\sqrt{(a_{5}-a_{1})^{2}-4a_{1}a_{8}}\right].
 \end{equation}
 The solution to the differential equation \eref{Gendiff} can be given by,
 \begin{equation}\label{gen-soln}
 \psi(x) = C_{\lambda_{\pm}}\sum_{m=0}^{\infty}(-1)^m\left[\frac{1}{(D+\lambda_{+})(D+\lambda_{-})}\{P_{+}+P_{-}\}\right]^{m}x^{-\lambda_{\pm}}.
 \end{equation} 
The above is exactly solvable only when $P_{-}=0$ and is quasi-exactly solvable under certain conditions, when both $P_{+}$ and $P_{-}$ are present \cite{Atre}.

\section{Exact and Quasi-Exact Solvability}\vspace{-1ex}
\label{sec2}
It has been shown that \cite{Turbiner}, quasi-exactly-solvable Schr\"{o}dinger equation that are connected to bilinear representation of $sl(2)$, 
can be cast in the differential form:
\begin{equation}\label{Turdiff}
-P_{4}(x)\frac{d^2\varphi}{dx^2}+P_{3}(x)\frac{d\varphi}{dx}+(P_{2}(x)-\varepsilon)\varphi = 0,
\end{equation}
where $P_{i}$'s are the polynomials of the $i$-th power $(i = 2,3,4)$ :
\begin{eqnarray*}
P_{4} = a_{++}x^4+a_{+0}x^3+(a_{+-}+a_{00})x^2+a_{0-}x+a_{--}, \\
P_{3} =
2(2j-1)a_{++}x^3+[(3j-1)a_{+0}+b_{+}]x^2 \\ +[2j(a_{+-}+a_{00})+a_{00}+b_{0}]x+ja_{0}+b_{-} ,\\
 -P_{2} = 2j(2j-1)a_{++}x^2+2j(ja_{+0}+b_{+})x+a_{00}j^2+b_{0}j
\end{eqnarray*}
We now investigate the conditions for exact and quasi-exact solvability of the aforementioned class of spectral problems.
Comparing this equation with \eref{Gendiff} yields $ a_{++}=a_{--}=0$ and leads to,
\begin{eqnarray*}
-a_{+0}=a_{0}, \ -(a_{+-}+a_{00})=a_{1}, \ -a_{0-}=a_{2}, \\ 2j(a_{+-}+a_{00})+a_{00}+b_{0}=a_{5}, \ ja_{0-}+b_{-}=a_{6}, \\ 
-2j(ja_{+0}+b_{+})=a_{7}, \ -(a_{00}j^2+b_{0}j)=a_{8}.
\end{eqnarray*}

The conditions that are to 
be imposed upon the differential equation \eref{Gendiff}, for exact-solvability and quasi-exact solvability are : 
\begin{equation}
\label{cond-exact}
a_{++} = a_{+0} = b_{+} = 0
\end{equation}
and
\begin{equation}
\label{cond-quasi-exact}
a_{0-} = a_{--} = b_{-} = 0,
\end{equation}
respectively.

 The exact solvability conditions \eref{cond-exact}, applied to \eref{Gendiff}, yield 
\begin{equation}\label{Hyper}
x\left(x+\frac{a_2}{a_1}\right)\frac{d^2\psi(x)}{dx^2}+\frac{a_5}{a_1}\left(x+\frac{a_6}{a_5}\right)\frac{d\psi(x)}{dx}+\frac{a_8}{a_1}\psi(x) = 0.
\end{equation}
 which is the \emph{Hypergeometric} differential equation. The two solutions can be 
 cast as a mapping connecting the monomial space to the space of polynomials \cite{Shreecharan} :
\begin{equation}
\psi_{\lambda_{\pm}}(x)=C_{\lambda_{\pm}}exp\left[\frac{1}{(D+\lambda_{\mp})}\left(a_{2}x\frac{d^2}{dx^2}+a_{6}\frac{d}{dx}\right)\right]x^{-\lambda_{\pm}}.
\end{equation}

The quasi-exact solvability conditions given in Table 2, produces a differential equation of the \emph{Heun }type
\fl\begin{equation}\label{qes-heun}
x^2\left(x+\frac{a_1}{a_0}\right)\frac{d^2\psi(x)}{dx^2}+\frac{a_4}{a_0}x\left(x+\frac{a_5}{a_4}
 \right)\frac{d\psi(x)}{dx}+\frac{a_7}{a_0}\left(x+\frac{a_8}{a_7}\right)\psi(x)=0.
\end{equation}
The two solutions to the above differential equation are \cite{pkpheun} 
\begin{equation}
\psi_{\lambda{\pm}}(x)= C_{\lambda_{\pm}}exp\left[\frac{(-1)}{(D+\lambda_{\mp})}\left(a_{0}x^3
\frac{d^2}{dx^2}+
a_{4}x^2\frac{d}{dx}+a_{7}x\right) \right]x^{-\lambda_{\pm}}.
\end{equation}
Under certain conditions, these series can terminate yielding polynomial solutions, that leads to known QES systems \cite{Atre}.

\FloatBarrier
\begin{table}
\caption{\label{math-tab2}Quasi-exact solvability conditions}
\begin{tabular*}{\textwidth}{@{}l*{15}{@{\extracolsep{0pt plus
12pt}}l}}
\br
Condition&Result&Implication\\
\mr
$a_{0-}=0$&$a_{2}=0$&Linear term in $f_{1}(x)$ is zero\\
$b_{-}=0$&$a_{6}=0$&Constant term in $f_{2}(x)$ is zero\\
\br
\end{tabular*}
\end{table}
\FloatBarrier

\nopagebreak

We now deal with the Heun equation, not amenable to the $sl(2)$ structure. 
We answer Turbiner's question affirmatively, by exploiting the algebraic symmetry, through a physical example.

Let us consider the equation
\begin{equation}\label{transformed}
\left(a\frac{d^2}{d\sigma^2}+b\frac{d}{d\sigma}+c+{\nu_n}^2 \right)\psi_n = 0
\end{equation}
where
\begin{eqnarray*}
a = {(1-\sigma^2)}^2(\sigma^2+\epsilon^2),\qquad
b = \sigma (1-\sigma^2)(1-2\epsilon^2-3\sigma^2),\\
c = -1+2\epsilon^2+6\sigma^2(2-\epsilon^2)-15\sigma^4\qquad \mbox{and}\qquad
\nu^2 = 4(1+\epsilon^2)\ {\omega_n}^2/\mu^2.
\end{eqnarray*}

This equation arises  from a Schr\"{o}dinger like equation, when one performs stability analysis around a kink solution 
of the $\phi^6$ theory in one dimension \cite{LeeChrist}. By defining \ $\zeta = \sigma^2$, $\psi_n =(1-\zeta)^s f $ where 
$s = \left[1-\left(\frac{\omega_n}{\mu}\right)^2\right]^\frac{1}{2}$, \eref{transformed} can be cast into the canonical form of the Heun equation:
\begin{equation}\label{Heun}
\frac{d^2f}{d\zeta^2}+\left[\frac{\gamma}{\zeta}+
\frac{\delta}{(\zeta-1)}+\frac{\varepsilon}{(\zeta-a)}\right]\frac{df}{d\zeta}+
\frac{\alpha \beta \zeta-q}{\zeta(\zeta-1)(\zeta-a)}f = 0,
\end{equation}
where
\begin{eqnarray*} 
\gamma=\varepsilon=\frac{1}{2},\delta=1+2s,a=-\epsilon^2,\\
\alpha=\left(-\frac{5}{2}-s\right),\beta=\left(\frac{3}{2}-s\right),\\
\mbox{and} \qquad
q = -\frac{1}{4}\left[1-2\epsilon^2-4(1-s^2)(1+\epsilon^2)+2s\epsilon^2\right].
\end{eqnarray*} 
The Heun equation is known to arise in several other contexts of physical interest 
\cite{Jha,Zecca,Ciprian,Boonserm,Hortacsu}. 
Equation \eref{Heun} can be written in the form,
\begin{equation}\label{Hamiltonian}
(\hat{H}-q)f= \{P_{+} - aP_{-}+F(P_{0})\}f =0,
\end{equation}
where 
\begin{numparts}
\begin{eqnarray}
P_{+}=\zeta^3\frac{d^2}{d\zeta^2}+(\gamma+\delta+\varepsilon)\zeta^2\frac{d}{d\zeta}+
\alpha\beta\zeta,\\
P_{-}=\zeta\frac{d^2}{d\zeta^2}+\gamma\frac{d}{d\zeta},\\
P_{0}=\zeta\frac{d}{d\zeta}-j,\\
\mbox{and}\qquad
F(P_{0})=n_{2}P_{0}^2+n_{1}P_{0}+n_{0}.
\end{eqnarray}
\end{numparts}
Here,
\begin{eqnarray*}
n_{2}=-(a+1),\qquad
n_{1}=\{a-(\gamma(a+1)+a\delta+\varepsilon)+1\}-2j(a+1),\\
n_{0}=-(a+1)j^2+\{a-(\gamma(a+1)+a\delta+\varepsilon)+1\}j-q.
\end{eqnarray*}
Here ${P_{+}}$ and ${P_{-}}$ are raising and lowering operators in the space of monomials 
$\{1,\zeta,\zeta^{2} ,\ldots ,\zeta^{2j}\}$, where $N=(2j+1)$ is the dimension of the space.
The closed algebra is a \emph{cubic} deformation of $sl(2)$,
\begin{eqnarray}
\left[P_{+},P_{-}\right] = \alpha_{1}P_{0}^3+\beta_{1}P_{0}^2+\gamma_{1}P_{0}+\delta_{1} \ ,\qquad
\left[P_{0},P_{\pm}\right] = \pm P_{\pm}
\end{eqnarray}
with
\vspace{-1ex}
\begin{eqnarray*}
\alpha_1 = 4,\qquad
\beta_1= 3(2\gamma+\delta+\epsilon-2)+12j,\\
\gamma_1= \{2\alpha\beta-3\gamma+2+(2\gamma-1)(\gamma+\delta+\epsilon)\\+6(2\gamma+\delta+\epsilon-2)j+9 j^2\},\\
\delta_1= \alpha\beta\gamma+ \{2\alpha\beta-3\gamma+2+(2\gamma-1)(\gamma+\delta+\epsilon)\}j\\+3(2\gamma+\delta+\epsilon-2)j^2+4 j^3.
\end{eqnarray*}
We now assume a polynomial solution of \eref{Heun} of the form, 
\begin{equation*}
f(\zeta^l)= g_{1}\zeta^{l}+g_{2}\zeta^{2l}+g_{3}\zeta^{3l}+\ldots
\end{equation*}
where $l=1$ or $l=\frac{1}{2}$, correspond to the independent variable $\sigma^2$ and $\sigma$ respectively in \eref{transformed}. 
Thus, to obtain a polynomial solution of definite degree, we terminate the polynomial 
at $\zeta^{n-1}$ and impose $P_{+}\zeta^{n-1} = 0$. This leads to
\begin{equation}\label{cond-n}
s^2+(2n-1)s+n^2-n-\frac{15}{4}=0.
\end{equation}

The above restriction allows, $n = 2, s = 1/2$ and $n = 3/2, s = 1$, yielding,
\begin{equation} 
\psi(x)_{n=2}=(1-\zeta)^{1/2}f(\zeta) = \left(\frac{\frac{\epsilon^2+1}
{\epsilon^2}}{\frac{\epsilon^2+1}{\epsilon^2}+\mbox{sinh}^2(\frac{\mu x}{2})}\right)^{1/2}    
 \frac{\mbox{sinh}^2(\frac{\mu x}{2})}{\left(\frac{\epsilon^2+1}{\epsilon^2}+
 \mbox{sinh}^2(\frac{\mu x}{2})\right)}.
\end{equation}
and 
\begin{equation}\label{2nd-state}
\psi(x)_{n=3/2}=(1-\zeta)f(\zeta^{1/2}) 
\left(\frac{\frac{\epsilon^2+1}{\epsilon^2}}{\frac{\epsilon^2+1}{\epsilon^2}+
\mbox{sinh}^2(\frac{\mu x}{2})}\right)   
\frac{\mbox{sinh}(\frac{\mu x}{2})}{\left(\frac{\epsilon^2+1}{\epsilon^2}+
\mbox{sinh}^2(\frac{\mu x}{2})\right)^{1/2}}.
\end{equation}
respectively, with $\label{soliton}
\sigma(x)=
\sinh({\mu x}/{2})\left[ \frac{\epsilon^2+1}{\epsilon^2}+\sinh^2({\mu x}/{2})\right]^{-\frac{1}{2}}$.
We note that the above is the ground state, since $P_{-}f (\zeta^{1/2})= 0$, leads to 
\begin{equation}
\left(\zeta\frac{d^2}{d\zeta^2}+\frac{1}{2}\frac{d}{d\zeta}\right)f(\zeta^\frac{1}{2}) = 0
\end{equation}
with the solution
\begin{equation}
f(\zeta^\frac{1}{2}) = 2c_{1}\zeta^{\frac{1}{2}}+c_{2}.
\end{equation}
Comparing this with $(\hat{H}-E)f(\zeta^\frac{1}{2})=0$, one has $c_{2}=0$, leading to the state $\psi(x)_{n=3/2}$.

\ It is worth pointing out, that we limited \eref{Gendiff} to only 3 singularities, which led to an algebra with three generators $\{P_{+},P_{0},P_{-}\}$. 
 Extension of this algebra can provide new symmetry and more general structure.
Assuming $f_{i}(x)$  to be ($i=1,\ldots,3$),
\begin{equation*}
f_{1}(x) = \sum_{n=0}^{N}a_{n}x^{n} ,\qquad
f_{2}(x) = \sum_{n=0}^{N-1}b_{n}x^{n} , \qquad
f_{3}(x) = \sum_{n=0}^{N-2}c_{n}x^{n}
\end{equation*}
\vspace{-1ex}
we get operators of order $(N$-2) and the algebra extends to\linebreak
$\{P_{N-2},P_{N-1},\ldots,P_{0},P_{-1},P_{-2}\}$. Note that generalized and tri-confluent Heun differential equation can be treated as examples of 
this extension \cite{Genheun,Sadoga}. These algebraic structures need careful study, which is currently under investigation. 

In conclusion, we have investigated the algebraic structure of a wide class of differential equations connected with spectral problems. It is found that, these 
equations, naturally possess deformed $sl(2)$ as their underlying symmetry algebra. The hidden symmetry of Heun differential equation is shown to be the cubic 
deformation of $sl(2)$ algebra, whereas the confluent and bi-confluent Heun lead to quadratic deformation. It is explicitly shown that exactly solvable and 
QES systems emerge from our analysis in an unified manner. We also explicitly show the algebraic structure of a dynamical systems not connected with 
the $sl(2)$ structure. In future, we would like to extend this investigation to other equations like tri-confluent and generalized Heun and generalize the 
study of this algebraic structure to higher dimensional symmetries.

\ack{}
A. Sen thanks IISER- Kolkata for a post-doctoral fellowship, during which a significant part of the present work has been carried out.
\nopagebreak


\section*{References}
\begin{thebibliography}{}

\bibitem{Flugge}
Fl\"{u}gge S 1994 Practical quantum mechanics (Springer-Verlag) and references therein.

\bibitem{Ushveridge}
Ushveridze A G 1994 Quasi-exactly solvable models in quantum mechanics (Taylor \& Francis) and references therein.

\bibitem{Turbiner}
Turbiner A V 1988 Commun. Math. Phys. \textbf{118} 467.

\bibitem{Bender}
Bender C M and Dunne G V 1996 J. Math. Phys. \textbf{37}(1) 6. 

\bibitem{Atre}
Atre R and Panigrahi P K 2003  Phys. Lett. A \textbf{317} 46.


\bibitem{Ronveaux}
Ronveaux A (ed.) 1995 \textit{Heun's differential equations} (Oxford University Press).

\bibitem{AKhare}
Jatkar D P, Kumar C N and Khare A 1989 Phys. Lett. A \textbf{142} 200.


\bibitem{Sudarshan}

Bhattacharjie A and Sudarshan E C G 1962 IL Nuovo Cimento \textbf{25} N.4\\
Gangopadhyaya A, Panigrahi P K and Sukhatme U P 1994 Helv. Phys. Acta \textbf{67} 363 and references therein.


\bibitem{Shifman}
Shifman M A 1989 Int. J. Mod. Phys. A \textbf{4}(12) 2897.

\bibitem{Rocek}
Rocek M 1991 Phys. Lett. B \textbf{255} 554.




\bibitem{nonlinalgeb}
Lakshmanan M and Eswaran K 1975 J. Phys. A: Math. Gen. \textbf{8} 1658.\\
Higgs P W 1979 J. Phys. A: Math. Gen. \textbf{12} 309.

\bibitem{deformap}
Curtright T L and Zachos C 1990 K Phys. Lett. B \textbf{243}(3) 237.\\
Zhedanov A S 1992 Mod. Phys. Lett. A \textbf{07} 507.

\bibitem{deform}
Abdesselam B, Beckers J, Chakrabarti A and Debergh N 1996 J. Phys. A: Math. Gen. \textbf{29}(12) 3075.\\
Quesne C 1999 J. Phys. A: Math. Gen. \textbf{32}(38) 6705.\\


\bibitem{Bambah}
Sunilkumar V, Bambah B A, Jagannathan R, Panigrahi P K and Srinivasan V 2000 J. Opt. B: Quantum Semiclass. \textbf{2} 126.

\bibitem{Govindarajan}
Govindarajan T R, Padmanabhan P and Shreecharan T 2010 J. Phys. A: Math. Gen. \textbf{43}(20) 205203.



\bibitem{Lay}
Lay W, Bay K and Slavyanov S Y 1998 J. Phys. A: Math. Gen. \textbf{31}(42) 8521.

\bibitem{Ovsiyuk}
Ovsiyuk E , Amirfachrian M and Veko O 2011 NPCS \textbf{15}(2) 163.

\bibitem{Kazakov}
Kazakov A Y and Slavyanov S Y 1996 Methods Appl. Anal. \textbf{3} 447.

\bibitem{Jha}
Jha P K and Rostovstev Y V 2010 Phys. Rev. A \textbf{81} 033827.



\bibitem{PKP}

Gurappa N and Panigrahi P K 2000 Phys. Rev. B \textbf{62} 3\\
Gurappa N, Panigrahi P K and Sreecharan T 2003 J. Comput. Appl. Math \textbf{160} 103\\
Gurappa N, Jha P K and Panigrahi P K 2007 SIGMA \textbf{3} 057.

\bibitem{Shreecharan}
Shreecharan T, Panigrahi P K and Banerji J 2004 Phys. Rev. A \textbf{69} 012102.

\bibitem{pkpheun}
Gurappa N and Panigrahi P K 2004 J. Phys. A: Math. Gen. \textbf{37} L605.

\bibitem{LeeChrist}
Christ N H and Lee T D 1975 Phys. Rev. D \textbf{12} 1606.

\bibitem{Zecca}
Zecca A 2010 Avd. Studies Theor. Phys. \textbf{4}(8) 353.

\bibitem{Ciprian}
Dariescu C and Dariescu M A 2012 Mod. Phys. Lett. A \textbf{27}(32) 1250184.

\bibitem{Boonserm}
Boonserm P and Visser M 2008 Phys. Rev. D \textbf{78} 101502.

\bibitem{Hortacsu}
Horta\c{c}su M 2013 arXiv preprint arXiv:1101.0471v3.


\bibitem{Genheun}
Norbert M, Hounkonnou M N and Ronveaux A 2011 Commun. Math. Anal. \textbf{11}(1) 121.

\bibitem{Sadoga}
Hounkonnou M N, Ronveaux A and Sadoga K 2007 Applied Math. Comp. \textbf{189} 816. 







\nopagebreak

\end{thebibliography}
\end{document}